# This photograph has been altered: Testing the effectiveness of image forensic labeling on news image credibility

Cuihua Shen, University of California, Davis
Mona Kasra, University of Virginia
James O'Brien, University of California, Berkeley

## Article's lead

Despite the ubiquity and proliferation of images and videos in online news environments, much of the existing research on misinformation and its correction is solely focused on *textual* misinformation, and little is known about how ordinary users evaluate fake or manipulated images and the most effective ways to label and correct such falsities. We designed a visual forensic label of image authenticity, Picture-O-Meter, and tested the label's efficacy in relation to its source and placement in an experiment with 2440 participants. Our findings demonstrate that, despite human beings' general inability to detect manipulated images on their own, image forensic labels are an effective tool for counteracting visual misinformation.

## Research questions

- How do image forensic labels showing the integrity (or lack thereof) of news images influence Internet users' credibility evaluation of these images?
- How does the placement of image forensic labels affect users' credibility perceptions of images? Are labels more effective when they are shown directly alongside the image (concurrent labeling), or after the fact (post-exposure labeling)?
- Does the source of image forensic labels matter? Would people trust software-generated labels more so than expert- or crowd-generated labels?

## Essay summary

- In a series of online experiments, 2440 participants from Amazon's MTurk saw photoshopped news images and rated these images' credibility. These images depicted various socio-political issues and were accompanied by a brief caption on simulated web and social media platforms such as Facebook and Twitter.
- Most participants saw the image with a forensic label showing that the image was either "Altered" or "Un-altered." The purported source of the label was either "software," "experts," or "other people on the Internet." The forensic label was placed either together with the image (concurrent placement) or after the image was viewed (post-exposure placement).
- Forensic labels are effective: Participants who saw images with an "Altered" label rated the image less credible than those who saw an "Un-altered" label or no label at all.
- Participants with higher digital media literacy were more skeptical of image credibility; Images that aligned with participants' pre-existing issue attitude were more likely to be perceived as credible.

- We did not find a "continued influence effect" of visual misinformation: Placing the labels concurrently with the image was as effective as placing them shortly after showing the image. Source effects were also mixed: "software" influenced participants' credibility judgments more than "human" when placed after image exposure, and "human" was more influential than "software" when placed concurrently with the image.

## Argument & Implications

Despite the ubiquity of images and videos in today's online environments, much of the existing research on misinformation and its correction strategies is solely focused on textual misinformation. Even though some studies have focused on image-oriented social media platforms such as Instagram, they are primarily concerned with the veracity of textual information embedded in images, rather than the images themselves (Vraga, Kim, Cook, & Bode, 2020). Little is known about how ordinary users process and evaluate fake or manipulated images in online news environments, and strategies to effectively counteract such falsities. As digital image creation and manipulation technology becomes increasingly advanced and accessible, the potential harmful consequences of visual misinformation cannot be overstated or ignored.

We argue that users process and perceive visual misinformation fundamentally differently than text-based misinformation. Visuals are detected and processed very quickly by the human brain (Potter, Wyble, Hagmann, & McCourt, 2014). They are more easily remembered, shared, and are often more persuasive than words. According to Dual Coding Theory (Clark & Paivio, 1991), visual information is processed independently from verbal (textual) information. While the perceptual features of visual information are similar or analogical to the events they depict, the relationship between words and their meanings is arbitrary. Empirical research has shown that imagery along with textual information produced better recall and memory of the message than the text alone (Clark & Paivio, 1991; Paivio & Csapo, 1973). On social media, news posts containing images capture individuals' attention more than those without any visuals (Keib et al., 2018), prompting more clicks and more user engagement (Li & Xie, 2020). Posts containing images are also more likely to go viral (Heimbach, Schiller, Strufe, & Hinz, 2015). Further, empirical evidence suggests that visual portrayals in news reporting have powerful framing effects, influencing audience perception of political figures (Peng, 2018) as well as opinions and behavioral intentions regarding political issues, much more so than text alone (Powell, Boomgaarden, De Swert, & de Vreese, 2015). Taken together, visual information's superior impression, retention, virality and persuasiveness all warrant special consideration from misinformation researchers.

This study addresses this critical gap. It demonstrates that image forensic labels are an effective means of correcting visual misinformation. Inspired by the Truth-O-Meter from Politifact (www.politifact.com), a widely-used fact checking website, we developed a visual barometer conveying forensic analysis of image integrity and veracity. We then tested the barometer's efficacy in relation to the source and placement of the forensic analysis in a series of online experiments with 2440 participants on MTurk. Our experiment reaffirmed previous findings that most human beings are unable to detect manipulated images on their own (Shen et al., 2019). Yet, carefully designed forensic labels can be a promising correction strategy in our battle against visual misinformation.

Two features of our study are noteworthy. First, it focuses solely on photographic images typically presented in a news context, while excluding explicitly synthetic images such as cartoons, infographics, or internet memes. This is because, unlike text, photographic images are believed to capture unedited truth. By contrast, synthetic images are more akin to text - they are expected to be created and heavily edited by designers and journalists. Second, this study is the first to test the efficacy of image forensic labels, rather than fact-checking labels. Most misinformation research to date focuses on the latter, which assesses the veracity of a claim and provides factual information should the claim be false. Forensic analysis of images assesses the integrity or veracity of the image itself. It evaluates whether the photographic image itself is altered or manipulated, regardless of the veracity of textual information embedded in or accompanying the image in the news context. Therefore, the well-studied fact-checking strategies in past research, such as logic and humor, no longer apply to image forensic analysis, which relies on techniques and insights much less accessible to the average user (Kasra, Shen, & O'Brien, 2018). The do's and don'ts for presenting image forensic analysis in online news contexts need to be systematically tested.

The most important takeaway from this study is that image forensic labels, in the form of a simple Picture-O-Meter, are effective in influencing credibility perception of images. Specifically, participants who saw an "Altered" label perceived the image to be less credible than those who saw either an "Un-altered" label or no label for the same image. In other words, people assume news images capture unedited truth, and they have to be persuaded otherwise. This finding is robust across various image content, news source and virality metrics. Consistent with past research (Kasra et al., 2018; Nightingale, Wade, & Watson, 2017), our study again suggests that human beings are incapable of detecting doctored images. Yet, a simple, low-tech forensic label like the Picture-O-Meter could make a considerable difference in our battle against misinformation. The forensic labels we tested in our experiment are easy and cheap to implement, provided the relevant authenticity determination is available. Unlike the prevalent fact-checking strategies used primarily for correcting textual misinformation, our labels were offered *without* any additional reason or explanation (other than that the image was considered "Altered" or "Un-altered" by either experts, other people on the internet, or software). We do not yet know if more explanations about how these forensic labels are derived would further enhance the labels' efficacy, and future studies should investigate this possibility. Still, it is encouraging news that forensic labels are a useful and feasible tool against visual misinformation for news and social media platforms.

Our finding further supports the critical importance of digital media literacy in the fight against misinformation. Participants with higher digital media literacy in general and digital photography experience specifically were more skeptical of image credibility. This means that interventions to boost digital media literacy could improve discernment of visual misinformation. There is some initial evidence showing the efficacy of such interventions (Guess et al., 2020), although these interventions are primarily focused on textual misinformation at the moment. Scholars and platforms should invest in developing programs and interventions that aim to improve visual digital literacy. We also found that images that align with participants' pre-existing issue attitude were more likely to be perceived as credible, consistent with "confirmation bias" found in prior research (Knobloch-Westerwick, Johnson, & Westerwick, 2015; Shen et al., 2019). Taken together, these findings suggest considerable individual differences in their susceptibility to visual misinformation, some of which can be potentially mitigated through media literacy interventions, while others are more enduring and issue-specific. This suggests that public agencies, educators, and platforms can use media literacy and issue-attitude to identify individuals susceptible to specific kinds of visual misinformation. Such susceptibility profiles can be created using a combination of self-reports and digital footprints such as browse and share history. These profiles can then be used for targeted inoculation and correction efforts.

We did not find a "continued influence effect," contradicting some existing research that misinformation perceptions may persist despite correction (Lewandowsky, Ecker, Seifert, Schwarz, & Cook, 2012; Walter & Tukachinsky, 2019). This could be partially attributed to our research design, as participants in the "post-exposure" condition saw the forensic label only a few minutes after seeing the image, so the misbelief may not be strong enough to linger. Source effects were also mixed: "software" worked better than "human" (either "experts" or "other people online") when placed after image exposure, and "human" worked better than "software" when placed concurrently with the image. Taken together, our results represent an initial step in understanding what roles placement and source of the forensic labels may play in counteracting visual misinformation, but significant nuances remain to be uncovered in future research.

In sum, we propose the following recommendations:
- Media organizations and platforms should develop and adopt forensic labels attached to news images to combat visual misinformation. These forensic labels are simple, brief, and easy to implement, and are effective on their own without providing further explanations.
- Media organizations and platforms should invest in interventions to boost digital media literacy, especially visual literacy. They could identify individuals susceptible to specific misinformation content and design targeted prevention and correction efforts.
- We call for more research to examine the placement and source effects on the efficacy of image forensic labels. Due to visual's unique properties, visual misinformation and its correction strategies need to be systematically tested.

## Findings

*Finding 1: Image forensic labels are effective - Participants who saw either no label or an "Un-altered" label rated the image more credible than those who saw an "Altered" label for the same image.*

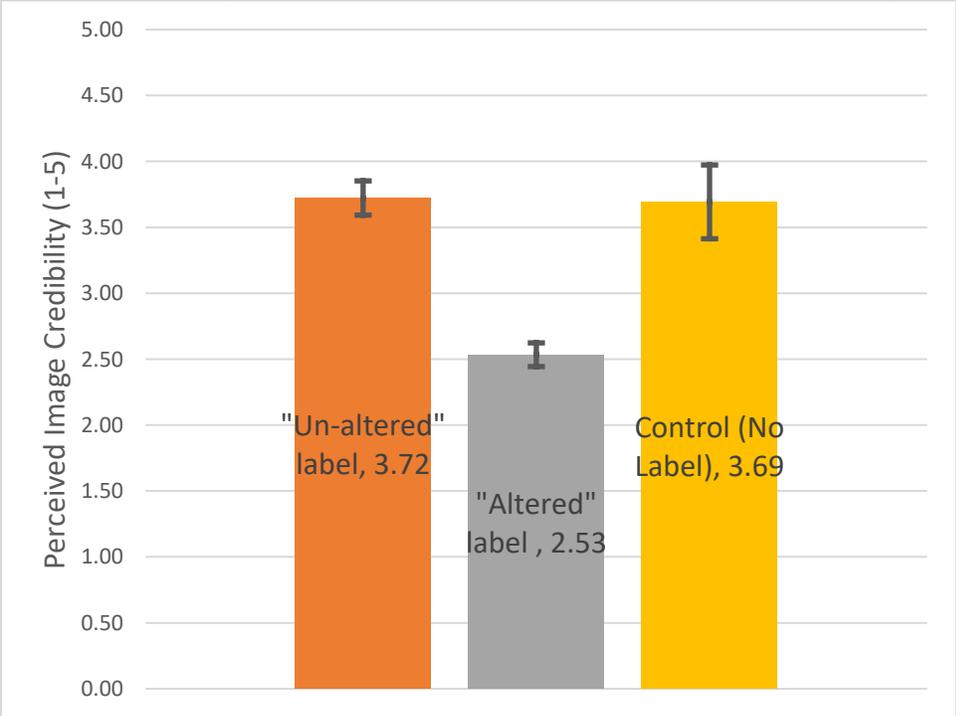

Figure 1. *Average image credibility rating by labeling condition. The "Altered" label condition was significantly lower than the control condition, while the "Un-altered" label condition did not differ significantly from the control condition. Error bars represent 95% confidence intervals for the results.*

We found that image forensic labels are a highly effective tool in swaying participants' credibility evaluation of images. Participants who were exposed to a forensic label showing an image was "Altered" rated the image 1.16 points less credible than those who did not see a label. By contrast, participants who viewed a forensic label of "Un-altered" rated the image just as credible as those who did not see a label. Our finding is highly robust across different images and credibility cues (e.g., number of likes and shares, whether the source was reputable such as New York Times), and after controlling for covariates in a follow-up analysis of covariance (see Appendix).

Our finding suggests that by default, people assume images are credible in online news environments, and they have to be persuaded otherwise. A simple forensic label like the Picture-O-Meter could make a considerable difference in credibility evaluation.

*Finding 2: a) Participants with higher digital media literacy were more skeptical of image credibility; b) Images that align with participants' pre-existing issue attitude were more likely to be perceived as credible.*

We found that image credibility evaluations differed considerably with participants' individual characteristics. People's prior experience with digital imaging and photography has a significant and negative association with credibility ratings (B = -0.15, p=.005), so did people's internet skills (B= -0.14, p=.005). Their pre-existing attitude supporting the issue depicted in the image showed a significant and positive association with credibility (B=0.20, p<.001). Participant's pre-existing political affiliation also mattered. People who self-identified as more liberal on the political spectrum tended to rate image credibility lower, although that association was marginally significant (B=- 0.05, p=.06). Participants' self-reported age and gender did not associate with how they rated credibility of these images.

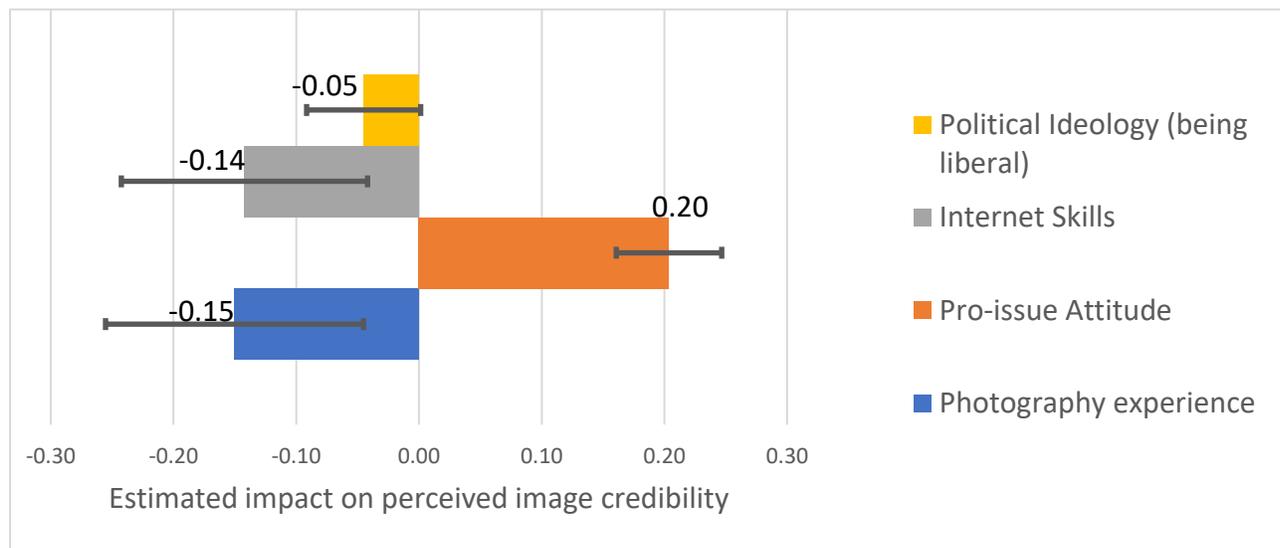

*Figure 2. Factors associated with participants' perceived image credibility rating. Numbers represent unstandardized regression coefficients. All factors were significant except political ideology, which was marginally significant (p=.06). Error bars represent 95% confidence intervals for the results.*

*Finding 3: We did not find a "continued influence effect" of visual misinformation: Placing the labels concurrently with the image was as effective as placing them after showing the image.*

Participants assigned to the post-exposure condition saw and rated the image credibility first, then were shown the forensic label, and rated the image again. Their second rating was significantly higher than the first rating for those who saw the "Un-altered" label ($M_{difference}$ = 0.33, $t$ =6.88, $p$ <.001), and significantly lower than the first rating for those who saw the "altered" label ($M_{difference}$ = -0.83, $t$ =-14.98, $p$ <.001), suggesting that the label was effective. Their second rating of image credibility was statistically equivalent to those of the concurrent condition (Participants exposed to the "Altered" label: $F(1, 978)$ = 1.96, $p$ =.16; Participants exposed to the "Un-altered" label: $F(1, 1000)$ = 0.39, $p$ =.53)), suggesting an absence of the "continued influence effect.". In other words, perceptions of visual misinformation dissipated with the forensic label.

Taken together, our finding is in direct contrast to previous research showing that misinformation belief tends to linger despite correction efforts (Walter & Tukachinsky, 2019). The discrepancy might be due to 1) the nature of visual information, 2) that there was only one exposure of the visual misinformation before the forensic label, and 3) the relatively short time lag between participants' credibility ratings before and after seeing the forensic label. More research is needed to test whether there is continued influence of visual misinformation with repeated exposures and longer time lags between exposure and correction.

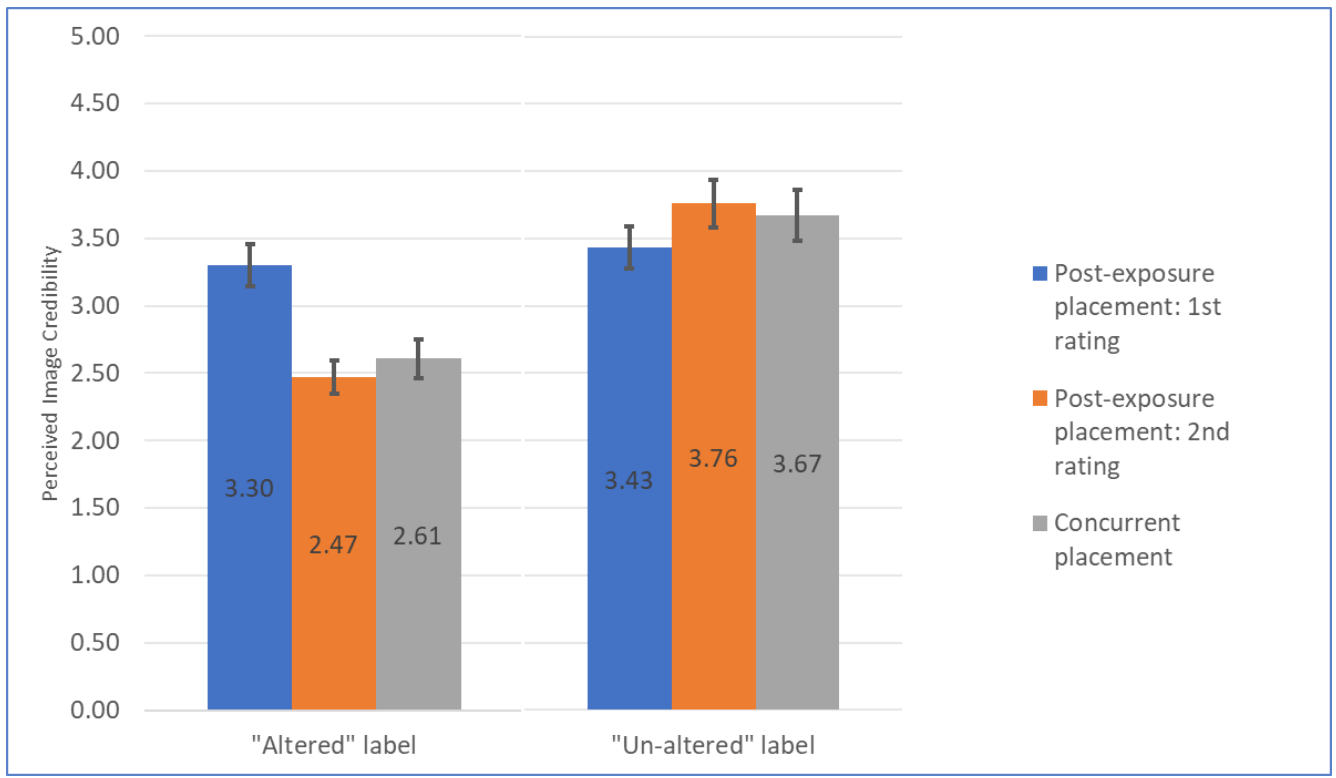

*Figure 3. Perceived image credibility in post-exposure condition (blue and orange bars) and concurrent condition (gray bars). Higher numbers represent greater perceived credibility. Paired-t tests showed that participants' second ratings (orange bars) were significantly different from their first ratings (blue bars), as a result of the forensic label, but they were not different from those of the concurrent condition (gray bars). Error bars represent 95% confidence intervals for the results.*

*Finding 4: As sources of image forensic labels, "software" worked better than "human" when placed after image exposure, and "human" worked better than "software" when placed concurrently with the image.*

The three sources of image forensic analysis (experts, other people online, and software) did not significantly differ from each other (see Appendix). After consolidating the "experts" and "other people online" categories into a "human" category, our exploratory analysis found significant interaction between forensic label source (label coming from either "human" or "software") and label placement (concurrent vs. post-exposure) (see Figure 4). Specifically, if the "Altered" forensic label came from software instead of humans (either experts or other people online), the label is more effective in reducing perceived credibility in post-exposure condition than in concurrent exposure condition. Similarly, if the "Un-altered" label came from software instead of humans, it was more effective in increasing perceived credibility in post-exposure condition, as compared to the concurrent exposure condition. In other words, "software" as a labeling source seems to amplify the label's effect if placed after people seeing the image, while "humans" as a labeling source was more effective when placed concurrently with the image.

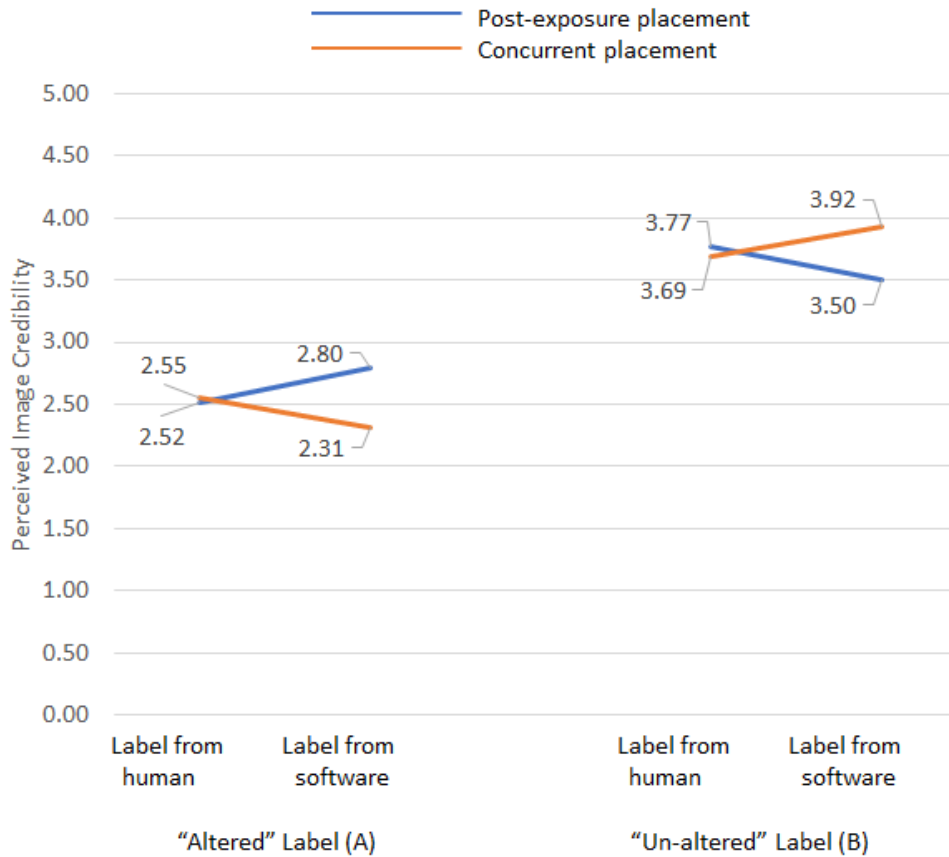

*Figure 4. Significant interactions between forensic label placement and source of forensic label. Higher numbers represent greater perceived image credibility. Panel (A) shows the participants exposed to the "Altered" label, and Panel (B) shows those exposed to the "Un-altered" label. The "Human" condition combines "Experts'' and "Other people online" conditions (see Appendix).*

## Methods

We designed a partial factorial experiment with 3 factors: Image forensic label ("Un-altered" / "Altered" / No label) X Source of forensic label (Expert / Other people online / Software) X Placement of forensic label (post-exposure/concurrent exposure/control- no forensic analysis), resulting in 13 unique conditions for each image tested (see Table 1).

Additionally, we included two more factors, image credibility cues (high/low) and image content (three unique images), which were manipulated across all conditions, bringing the total number of experimental cells to 78 (see Appendix). For each image, we included a one-line brief textual caption to simulate how people typically consume images in online news environments (see Figure 5 for an example stimulus). To make sure no participants had prior exposure to our stimuli, all three images and their accompanying captions used were fake - they were purposefully cropped, changed and combined (for more information on stimuli creation, see Shen et al 2019).

Table 1. Partial Factorial Experimental design with three factors: Image forensic label, Source of forensic analysis, and Placement of forensic label.

| Condition | Placement of Forensic Label | Forensic Label | Source of Label |
|---|---|---|---|
| 1 | Post exposure | Altered | Experts |
| 2 | Post exposure | Altered | Other People |
| 3 | Post exposure | Altered | Software |
| 4 | Post exposure | Un-altered | Experts |
| 5 | Post exposure | Un-altered | Other People |
| 6 | Post exposure | Un-altered | Software |
| 7 | Concurrent exposure | Altered | Experts |
| 8 | Concurrent exposure | Altered | Other People |
| 9 | Concurrent exposure | Altered | Software |
| 10 | Concurrent exposure | Un-altered | Experts |
| 11 | Concurrent exposure | Un-altered | Other People |
| 12 | Concurrent exposure | Un-altered | Software |
| 13 | Control (no forensic analysis) | | |

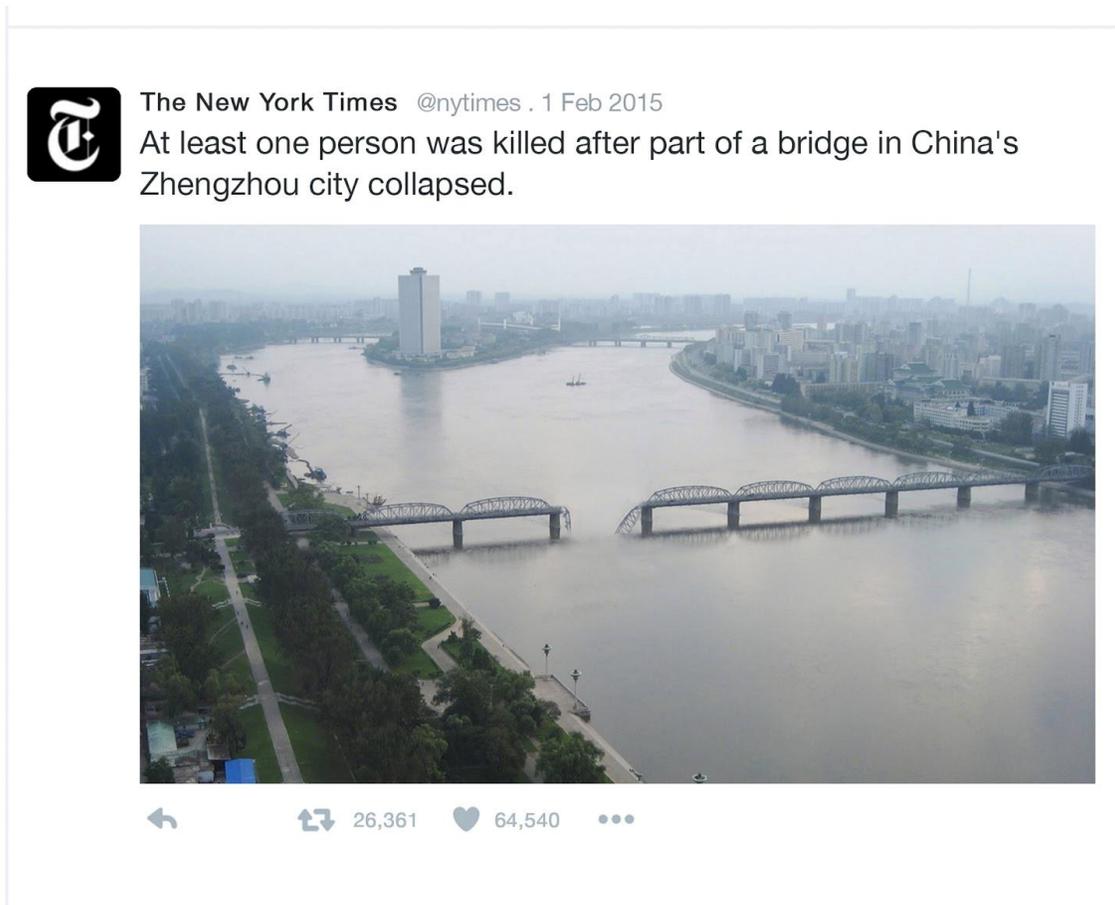

*Figure 5. An example news image tested in the study*

**Image forensic label.** Inspired by the fact-checking website Politifact's veracity barometer, "Truth-O-Meter," we designed a visually similar barometer (Picture-O-Meter) to convey image forensic information. To make the barometer as unambiguous as possible, there were only two labels on the barometer, "Un-altered" and "Altered" (see Figure 6), with a short description accompanying the barometer: "UN-ALTERED - The picture is original and untouched" or "ALTERED - The picture is fake and manipulated."  A qualitative user assessment with a few college undergraduates showed that Picture-O-Meter was clear and easy to understand.

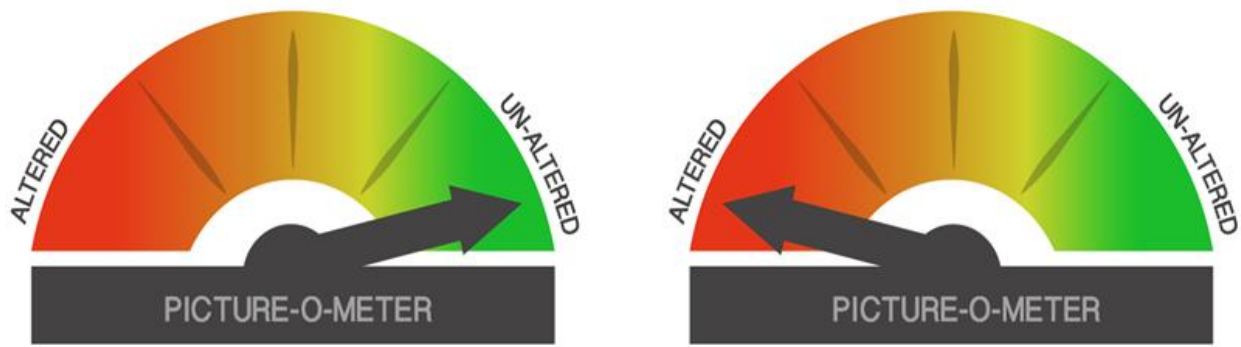

*Figure 6. Picture-O-Meter showing the "Un-Altered" and "Altered" forensic labels.*

**Source of forensic analysis**. We manipulated the purported source of image forensic analysis by presenting a short description alongside the Picture-O-Meter: "The Picture-O-Meter indicates the level of authenticity of the photo as rated by **experts on detecting photo editing and manipulation (software for detecting photo editing and manipulation; other people on the internet)** " (emphasis original). The "experts" and "other people" categories were later merged into a "human" category in analysis (see Appendix).

**Placement of forensic analysis**.  In the post-exposure placement condition, participants saw and rated the image first without any forensic analysis information (first rating), then they answered other questions before seeing the same image accompanied by forensic analysis shown in the picture-o-meter on the same screen, and rated its credibility again (second rating). These two credibility tests were kept as distant from each other as possible in the survey experiment. In the concurrent placement condition, the forensic label was shown on the same screen as the image itself. Participants only had one opportunity to evaluate the credibility of the image. In the control condition, participants saw the image once and rated its credibility, without any forensic label.

Participants were recruited from Mechanical Turk ([www.mturk.com](www.mturk.com)) and were redirected to Qualtrics to complete a survey study. The experiment was conducted in three batches consecutively, each featuring one image, within a 6 week period in 2017. The control condition (Condition 13, N=157) was taken from the user sample of a previous study (Shen et al., 2019), while participants in Conditions 1 - 12 were recruited anew. Participants were only exposed to one image in the study, and could complete the study only once.

Participants first read the consent form and confirmed to proceed to one of the twelve randomly assigned experimental conditions (Conditions 1-12) on Qualtrics. Since some questions required knowledge of U.S. politics, participants who were younger than 18 or resided outside of the United

States were excluded after the initial screening questions. To ensure consistent image dimensions, we instructed participants on a mobile device to switch to a laptop or desktop computer before proceeding to the questions or they would be excluded from the study.

Participants first answered questions about their internet skills and political attitude, then evaluated the credibility of an image, followed by questions on demographics. To make sure enough attention was paid to the image itself, participants could not proceed until 30 seconds had lapsed on that screen. For those who were assigned to the post-exposure label condition and had to evaluate the credibility of the same image twice, these two evaluations were placed in the survey as distant from each other as possible. The survey study took about 5 minutes to complete, and all participants were paid $0.25 for the task.

2814 participants from MTurk attempted to participate in the study, among which 2440 completed the main questions (86.71%). There were 1817 (74.5%) participants identified as White/Caucasian, 225 (9.2%) as African American, 150 (6.1%) as Hispanic, 160 (6.6%) as Asian, 17 (0.7%) as Native American, 9 (0.4%) as Pacific Islander, 46 (1.9%) as other race, and 16 (0.7%) who did not disclose their race. 1065 (43.6%) were identified as men, 1363 (55.9%) as women, and 12 (0.5%) did not close their gender. The median age was 27 years old (Mean = 29.72). 13 (0.5%) had less than high school education, 243 (10.0%) had high school or GED, 618 (25.3%) had some college, 274 (11.2%) had 2-year college degree, 896 (36.4%) had 4-year college degree, 15.2% had doctoral or professional degree. The median income category was $50,000-$59,999. Compared to the 2019 US Census, the study sample had slightly more women, white/Caucasian and Asian than in the general US population. For detailed information on measures and analysis, please refer to Appendix.

## Bibliography

Clark, J. M., & Paivio, A. (1991). Dual Coding Theory and Education. *Educational Psychology Review, 3*(3), 149-210.
Guess, A. M., Lerner, M., Lyons, B., Montgomery, J. M., Nyhan, B., Reifler, J., & Sircar, N. (2020). A digital media literacy intervention increases discernment between mainstream and false news in the United States and India. *Proceedings of the National Academy of Sciences, 117*(27), 15536. doi: 10.1073/pnas.1920498117
Heimbach, I., Schiller, B., Strufe, T., & Hinz, O. (2015). *Content Virality on Online Social Networks: Empirical Evidence from Twitter, Facebook, and Google+ on German News Websites*. Paper presented at the Proceedings of the 26th ACM Conference on Hypertext & Social Media, Guzelyurt, Northern Cyprus. https://doi.org/10.1145/2700171.2791032
Kasra, M., Shen, C., & O'Brien, J. F. (2018). *Seeing Is Believing: How People Fail to Identify Fake Images on the Web*. Paper presented at the Extended Abstracts of the 2018 CHI Conference on Human Factors in Computing Systems, Montreal QC, Canada.
Keib, K., Espina, C., Lee, Y.-I., Wojdynski, B. W., Choi, D., & Bang, H. (2018). Picture This: The Influence of Emotionally Valenced Images, On Attention, Selection, and Sharing of Social Media News. *Media Psychology, 21*(2), 202-221. doi: 10.1080/15213269.2017.1378108
Knobloch-Westerwick, S., Johnson, B. K., & Westerwick, A. (2015). Confirmation Bias in Online Searches: Impacts of Selective Exposure Before an Election on Political Attitude Strength and Shifts. *Journal of Computer-Mediated Communication, 20*(2), 171-187. doi: 10.1111/jcc4.12105

Lewandowsky, S., Ecker, U. K. H., Seifert, C. M., Schwarz, N., & Cook, J. (2012). Misinformation and Its Correction: Continued Influence and Successful Debiasing. *Psychological Science in the Public Interest, 13*(3), 106-131. doi: 10.1177/1529100612451018

Li, Y., & Xie, Y. (2020). Is a Picture Worth a Thousand Words? An Empirical Study of Image Content and Social Media Engagement. *Journal of Marketing Research, 57*(1), 1-19. doi: 10.1177/0022243719881113

Nightingale, S. J., Wade, K. A., & Watson, D. G. (2017). Can people identify original and manipulated photos of real-world scenes? *Cognitive Research: Principles and Implications, 2*(1), 30. doi: 10.1186/s41235-017-0067-2

Paivio, A., & Csapo, K. (1973). Picture superiority in free recall: Imagery or dual coding? *Cognitive Psychology, 5*(2), 176-206. doi: https://doi.org/10.1016/0010-0285(73)90032-7

Peng, Y. (2018). Same Candidates, Different Faces: Uncovering Media Bias in Visual Portrayals of Presidential Candidates with Computer Vision. *Journal of Communication, 68*(5), 920-941. doi: 10.1093/joc/jqy041

Potter, M. C., Wyble, B., Hagmann, C. E., & McCourt, E. S. (2014). Detecting meaning in RSVP at 13 ms per picture. *Attention, Perception, & Psychophysics, 76*(2), 270-279. doi: 10.3758/s13414-013-0605-z

Powell, T. E., Boomgaarden, H. G., De Swert, K., & de Vreese, C. H. (2015). A Clearer Picture: The Contribution of Visuals and Text to Framing Effects. *Journal of Communication, 65*(6), 997-1017. doi: 10.1111/jcom.12184

Shen, C., Kasra, M., Pan, W., Bassett, G. A., Malloch, Y., & O'Brien, J. F. (2019). Fake images: The effects of source, intermediary, and digital media literacy on contextual assessment of image credibility online. *New Media & Society, 21*(2), 438-463. doi: 10.1177/1461444818799526

Vraga, E. K., Kim, S. C., Cook, J., & Bode, L. (2020). Testing the Effectiveness of Correction Placement and Type on Instagram. *The International Journal of Press/Politics, 25*(4), 632-652. doi: 10.1177/1940161220919082

Walter, N., & Tukachinsky, R. (2019). A Meta-Analytic Examination of the Continued Influence of Misinformation in the Face of Correction: How Powerful Is It, Why Does It Happen, and How to Stop It? *Communication Research, 47*(2), 155-177. doi: 10.1177/0093650219854600

# Appendix

*Experimental design: Additional factors*

Image credibility cues (not hypothesized). The "high credibility" condition was achieved in using New York Times as the purported source, with 64,540 favorites and 26,361 retweets as the bandwagon metrics. The "low credibility" condition was achieved by using a generic person's Twitter account (Rachael Hughes) as the purported source of the image, with only one favorite and one retweet as the bandwagon metrics. The image sources were selected based on a Pew report on media trustworthiness (Mitchell et al., 2014), which ranked New York Times as one of the most trustworthy news sources. Both the purported sources and virality metrics were validated and used in a previous study (Shen et al, 2019).

Image content (not hypothesized). These images were used in a previous study (Shen et al, 2019) and represented different sociopolitical issues with varied media exposure in recent years.

These two factors were included to expand ecological validity of the study, not to test their separate effects on the outcome variable, so they were manipulated but not explicitly tested in the analysis.

*Measures*

Perceived credibility. This variable was measured by six items of perceived credibility adapted from Flanagin and Metzger's (2007) on a 7-point scale (1=strongly disagree, 7= strongly agree). It assessed the extent to which participants perceived the image to be believable, original, authentic, fake, manipulated, and retouched. After reverse-coding negatively-worded items, the mean was taken to create a composite credibility score (Cronbach's alpha=.95). In the concurrent exposure and control conditions, credibility was measured only once. In the post-exposure condition, the perceived credibility was measured twice, once before seeing the barometer and once after. We also calculated the net credibility change by subtracting the pre-barometer rating from the post-barometer rating.

Internet skills. Participants' Internet skills were measured by their familiarity with 10 Internet-related terms (e.g. phishing and spyware) on a 5-point Likert scale (Hargittai & Hsieh, 2012). Then, the mean of these items became a composite score of Internet skills (Cronbach's alpha =.92).

Digital imaging skills. Two items were used to measure participants' photography and digital imaging (e.g. photo editing) experiences and skills (Greer & Gosen, 2002) on a 5-point scale (1=None, 5=I'm an expert). The mean was then taken to be the composite measure of digital imaging skills (Cronbach's alpha=.74)

Pro-issue attitude. For each of the three images tested, two items were used to measure participants' preexisting attitudes toward the issue depicted in the image. These items were adapted from Treier and Hillygus (2009) and modified to fit each of the images tested. For example, participants evaluating the image showing a genetically modified mouse were asked whether it is ethical or acceptable to genetically modify animals for research purposes. Negatively worded questions were reversed coded, and then the two items were averaged to create a composite score of pro-issue attitude (Cronbach's alpha=.81).

Political ideology. Participants were asked to indicate their political ideology on a 7-point likert scale, from extremely conservative (1) to extremely liberal (7).

Internet use. Participants were asked how many years they have been using the Internet, and also how many hours on average per day they use the Internet for non-work reasons.

Demographics. At the end of the survey, participants were asked to indicate their sex, age, race, annual household income, and education level. Participants' age and sex were included in our analysis as control variables.

*Manipulation check*

We performed a manipulation check of the forensic label by asking participants to indicate what forensic designation the barometer was pointing at (Un-altered, Altered, or not sure). Among the 2440 participants who completed the study, 2283 were exposed to an image forensic barometer (conditions 1-12), of which 1982 (86.8%) correctly recalled its forensic designation, and 301 (13.2%) either answered the wrong designation or "unsure.". As expected, those who failed the manipulation check rated the image more credible than those who identified the forensic designation correctly (Mean failed = 3.39, Mean passed = 3.13, t= -2.49, p=0.01). A chi-square test showed that participants in the post-exposure placement condition were more likely to fail the manipulation check than those assigned to the concurrent placement condition (chi-square = 37.335, df=1, p <.001). In the following analysis, these participants who failed the manipulation check (N=301) were excluded, leaving a final sample of 2139 participants.

*Finding 1 and Finding 2 analysis*

To test the main effect of image forensic labeling, we ran analyses in two stages. In the first stage, an omnibus ANOVA showed a significant main effect ($F(2, 2136)=112.38$, $p<.001$). Multiple comparisons using Dunnett T3 and Bonferroni adjustment showed that, participants exposed to the "Altered" label rated the image significantly less credible than those who did not see the label ($M_{diff}$ = -1.16, $p<.001$), but those exposed to the "Un-altered" label did not differ from the control group ($M_{diff}$=0.03, p=.99). In the second stage, we ran an analysis of covariance (ANCOVA) with perceived credibility of the image (the second credibility rating for post-exposure condition) as the dependent variable, and image forensic label as the main factor, while also including the respondent's age, gender, political ideology, and issue attitude as covariates. Results still showed a significant main effect of image forensic label ($F(2, 2126)=120.96$, $p<.001$). A planned contrast between the "Un-altered" condition and the control condition showed a non-significant difference ($M_{diff}$ = -0.129, SE = 0.15, p=.40), while participants in the the "Altered" condition rated image significantly less credible than those in the control condition (diff = -1.31, SE=0.15, $p<.001$). Therefore, both ANOVA and ANCOVA showed the same results.

Among the covariates, people's prior experience with digital imaging and photography has a significant and negative association with credibility ratings ($F(1, 2126)=7.86$, p=.005, B = -0.15), so did people's internet skills ($F(1, 2126)=7.72$, p=.005, B= -0.14). Their pre-existing attitude supporting the issue depicted in the image showed a significant and positive association with credibility ($F(1, 2126)=86.45$, $P<.001$, B=0.20). Participant's pre-existing political affiliation (F=3.62, p=.06), age ($F(1, 2126)=0.93$, p=.34), and gender ($F(1, 2326)=0.58$, p=.45) did not associate with how they rated credibility of these images.

To probe whether the results differed across the two image credibility cues conditions (high vs. low credibility cues), we ran a post-hoc ANCOVA with image credibility cues as an additional factor, along with its interaction term with image forensic designation. We found that both the main effect of credibility cues ($F(1, 2123)=2.20$, p=.14) and the interaction between credibility cues and forensic labels ($F(1, 2123) = 1.446$, p=.229) were nonsignificant. Therefore, our results are robust across different credibility cue manipulations.

*Finding 3 analysis*

In order to test whether participants' exposure to visual misinformation would have a continued influence effect after they were shown the forensic label, we ran paired sample t tests between their first credibility rating (before seeing the forensic label) and their second credibility rating of the same image (after seeing the forensic label). Their second rating was significantly higher than the first rating for those who saw the "Un-altered" label ($M_{difference}$ = 0.33, t=6.88, $p<.001$), and significantly lower than the first rating for those who saw the "Altered" label ($M_{difference}$ = -0.83, t=-14.98, $p<.001$). Additionally, ANOVA tests showed that participants' second rating of image credibility and was statistically equivalent to those of the concurrent condition (Participants exposed to the "Altered" label: $F(1, 978)=1.96$, p=.16;

Participants exposed to the "Un-altered" label: F(1, 1000)=0.39, p=.53). To test whether the results were robust across different image credibility cue conditions and with covariates, we ran ANCOVA models with image credibility cues as an additional factor. Results were virtually unchanged, and no significant difference was found across the high and low credibility cue conditions.

*Finding 4 analysis*

To test the main effects of labeling source, we ran two sets of models, one with the participants who were shown the "Altered" label, and the other with participants shown the "Un-altered" label. The omnibus ANOVA test showed that the source of forensic label with three levels (experts, other people online, and software) on its own was not associated with participants' credibility perception of the images (Participants exposed to the "Altered" label: F(2, 977)=2.25, p=.11; Participants exposed to the "Unaltered" label: F(2, 999)=0.44, p=.64).

Post-hoc two-way ANOVA of both source and placement of forensic labels showed that the interaction between rating source and placement was significant for participants exposed to the "Altered" label (F(1, 974)=3.31, p=.04) but not for participants exposed to "Un-altered" label (F(2, 996)=1.70, p=.18). Specifically, if the "Altered" label's forensic analysis purportedly came from software instead of experts or other people online, its association with people's credibility perception bifurcated in prepost and concurrent conditions (see Figure A1).

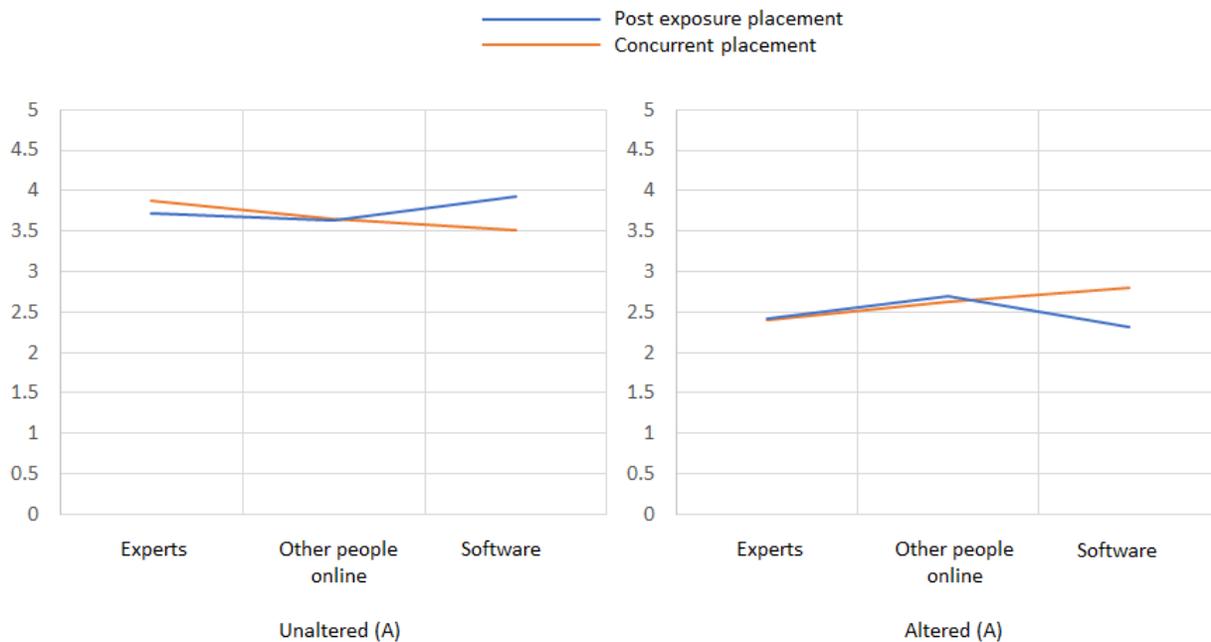

*Figure A1. Interactions between forensic label placement and source of forensic label. Higher numbers represent greater perceived image credibility. Panel (A) shows the participants exposed to the "Un-altered" label, and Panel (B) shows those exposed to the "Altered" label.*

Based on the above results, we consolidated the source of forensic analysis into just two levels: "human" (combining "expert" and "other people online") versus "software." We then ran two-way ANOVA with the source and placement of forensic labels. Results again showed a significant interaction for those seeing the "Altered" label (F(1, 976)=6.67, p=.01), and a marginally significant interaction for those seeing the "Un-altered" label (F(1, 998)=3.25, p=.07).

To test whether the results differed across the two image credibility cues conditions (high vs. low credibility cues), we further added image credibility cues as another factor along with covariates. Main results are unchanged from previous models, and the three-way interaction among credibility cues, rating source and placement of forensic analysis was not significant (participants seeing the "Altered" label: F(1,965)=0.29, p=.59; participants seeing the "Un-altered" label: F(1,985)=0.60, p=.43), showing that the results were robust and did not differ across high and low image credibility cue conditions.

*Bibliography*

Flanagin, A. J., & Metzger, M. J. (2007). The role of site features, user attributes, and information verification behaviors on the perceived credibility of web-based information. New Media & Society, 9(2), 319-342. doi: 10.1177/1461444807075015

Greer, J. D., & Gosen, J. D. (2002). How much is too much? Visual Communication Quarterly, 9(3), 4-13. doi: 10.1080/15551390209363485

Hargittai, E., & Hsieh, Y. P. (2012). Succinct Survey Measures of Web-Use Skills. Social Science Computer Review, 30(1), 95-107. doi: 10.1177/0894439310397146

Mitchell, A., Gottfried, J., Kiley, J., & Masta, K. E. (2014). Political Polarization & Media Habits. Retrieved Oct 31, 2017, from http://www.journalism.org/2014/10/21/political-polarization-media-habits/#trust-and-distrust-liberals-trust-many-conservatives-trust-few

Shen, C., Kasra, M., Pan, W., Bassett, G. A., Malloch, Y., & O'Brien, J. F. (2019). Fake images: The effects of source, intermediary, and digital media literacy on contextual assessment of image credibility online. New Media & Society, 21(2), 438-463. doi: 10.1177/1461444818799526

Treier, S., & Hillygus, D. S. (2009). The Nature of Political Ideology in the Contemporary Electorate. The Public Opinion Quarterly, 73(4), 679-703.